**Intrinsic deep hole trap levels in $Cu_2O$ with self-consistent repulsive Coulomb energy**

Bolong Huang*
*Department of Physics and Materials Science, City University of Hong Kong, Kowloon, Hong Kong SAR, China*
*Present Address: Department of Applied Biology and Chemical Technology, The Hong Kong Polytechnic University, Hung Hom, Kowloon, Hong Kong SAR, China*

Email: bhuang@polyu.edu.hk

The large error of the DFT+U method on full-filled shell metal oxides is due to the residue of self-energy from the localized d orbitals of cations and p orbitals of the anions. U parameters are self-consistently found to achieve the analytical self-energy cancellation. The improved band structures based on relaxed lattices of $Cu_2O$ are shown based on minimization of self-energy error. The experimentally reported intrinsic p-type trap levels are contributed by both Cu-vacancy and the O-interstitial defects in $Cu_2O$. The latter defect has the lowest formation energy but contributes a deep hole trap level while the Cu-vacancy has higher energy cost but acting as a shallow acceptor. Both present single-particle levels spread over nearby the valence band edge, consistent to the trend of defects transition levels. By this calculation approach, we also elucidated the entanglement of strong *p-d* orbital coupling to unravel the screened Coulomb potential of fully filled shells.

Cuprous oxide ($Cu_2O$) is direct gap semiconducting oxide with full/nearly filled shell for Cu. It has been regarded as one of the most promising candidate of photovoltaic cells. It is also a prototype materials of "invisible electronic devices" derived from $CuMO_2$ (M=Al, Ga, In, Cr, etc.) with wide energy range of p-type doping limit determined by intrinsic defects[1]. Using the DFT+U method on these materials, we can rapidly determine electronic structures of the atomic models[2]. Furthermore, a linear response method can be used to obtain a localized partially filled model; however, it is difficult to obtain the correct U parameter estimate for the $3d^{10}$ configuration. The electron wavefunctions of $3d^{10}$ are constrained with strict boundary conditions. Therefore, the perturbation becomes extremely small if a small Lagrange multiplier is used to perturb the fully filled localized orbital. The inverse of this small difference tends to be a singularity (e.g., $1/\chi$ with $\chi \to 0$). Thus, the simple U parameter estimation through the small perturbation inverse using linear response becomes unphysical. Therefore, the first issue is start from the $3d^{10}$ orbital energy.

More unusually, an interesting scenario happens in ZnO also with $3d^{10}$ configuration for $Zn^{2+}$. The electronic structures and lattice relaxation exhibit a strong correlation for zincblend, rocksalt, and wurtzite phases of ZnO. Ma et al has empirically tuned U parameters for 3d orbitals of Zn and 2p orbitals of O respectively, and shown this effect[3] in terms of band structure, lattice geometry, and native defect levels. And this effect occurs regardless the local atomic coordination, which hints an intrinsic feature. However, we need to understand the reason through the theory level of self-consistent determination *p-d* orbital entanglement.

On the other hand, the actual 3d orbital occupations between cubic $Cu_2O$ and monoclinic CuO in antiferromagnetic phase are still unknown by DFT+U. There has been a long debate between $3d^{10}$ and $3d^9$ for ground state $Cu_2O$ and CuO. The band structure calculations of $Cu_2O$ by DFT+U always underestimate the band gap due to strong *p-d* orbital entanglement. This directly leads to incorrect density of states for p and d orbitals known from Sieberer et al[4] where the p and d orbitals levels were shown to be mixed at valence band maximum. Robertson et al[5] has given accurate band structure and optical property of $Cu_2O$, and confirmed that the 2p orbital of O is in fact lower than the 3d orbital of Cu and 3d levels contribute the VBM. We can understand this with combined aspects of binding energy and structure symmetry. The crystal structure of $Cu_2O$ is simple cubic, $O_h^4$, but the Cu site has rather seen the linear O-Cu-O coordination as cation in oxides, while the O sites are tetrahedrally coordinated by Cu. This means the O site in $Cu_2O$ lattice has more coordination number than Cu site, leading to a stronger binding energy than the ordinary O sites in the lattice of other oxides, where the coordination number of O site is always lower than the one of nearest neighboring cations. More advance method like hybrid functional has shown coherent orbital energies of the band structure but left a potential error for band gap by Heinemann et al[6] and Robertson et al[7].

The estimation of the Coulomb repulsive potential in DFT+U is tried by Cococcioni et al[8, 9] but fails in predicting the energy for fully occupied orbitals. This arises because the pristine Janak theorem that linear response relies on omitted the spurious Coulomb self-energy of the semicore orbitals. This amplifies the error of self-energy term when applying U on such orbitals for projecting the semicore states out. Regarding the Coulomb self-energy correction of Perdew and Zunger[10], the strict condition of correction has been updated into Janak equation[11] as following forms.

$$\begin{cases} \left(\dfrac{\delta \vec{E}}{\delta n_i}\right)_{cation} = (\varepsilon_i)_{cation} + \sum_{cation} \\ \left(\dfrac{\delta \vec{E}}{\delta n_j}\right)_{anion} = (\varepsilon_j)_{anion} + \sum_{anion} \end{cases} \quad (1)$$

The *i* and *j* denotes the *i*th cation and *j*th anion respectively. The $(\varepsilon_i)_{cation}$ and $(\varepsilon_j)_{anion}$ are orbital eigenvalues for lining up the band structures. The $\sum_{cation}$ and $\sum_{anion}$ are self-energies induced by semicore states of cations and anions, to be annihilated ideally fowling the condition of $U[n_{\alpha\sigma}] + E_{XC}[n_{\alpha\sigma}, 0] = \sum = 0$, by Perdew and Zunger. By a self-consistent linear response procedure, the U parameters assigned to both cations and anions are reliably obtained[12]. FIG. 1 shows how the self-energy to be counteracted in fully occupied shell of cations. We choose the non-linear core-corrected norm-conserving pseudopotentials for both the cation and anion elements[2]. The norm-conserving pseudopotentials reflect the all-electron behavior for the outer shell valence electrons with **|S-matrix|**=1 compared with ultrasoft pseudopotentials[13, 14].

The origin of the intrinsic p-type conductivity in $Cu_2O$ has been under investigation for a long time. Experiment reported the two acceptor-like levels with two different trends of densities variation under increasing oxygen chemical potential[15]. Scanlon et al. proposed that the intrinsic p-type conductivity in $Cu_2O$ is attributed to Cu binding to Cu vacancy with tetrahedral coordinate ($V_{Cu}^{split}$) using a HSE study and contributes to a deep localized state [16, 17]. However, Isseroff et al. [18] used the same HSE method and determined that the $V_{Cu}^{split}$ is approximately 0.5 eV higher than normal $V_{Cu}$. This is attributed to oxygen hole levels that are not well counteracted. Both Raebiger et al [19] and Soon et al [20] reported the O-interstitials act as deep hole trap levels, but with higher formation energy than the Cu-vacancy which gives shallower hole trap levels. This requires to confirm by accurate band gap and oxygen hole levels.

However, recent experiment [15] shows that the concentration of deeper trap level is higher at high temperature (>750 K) or with relatively wide range of oxygen chemical potential.

It is more efficient to observe the charge transition level in the band gap, which denotes the defect or dopants thermal ionization energy. We call the energy level is the thermal transition energy level or thermodynamic transition level (TTL).

The thermal transition energy/level (TTL) $\varepsilon(q/q')$ is the critical Fermi level position in the band gap where the charge state changes from q to q' as $\Delta E_F$ changes in the band gap with the lowest-energy, which means the formation energy follows $\Delta H(q, E_F)=\Delta H(q', E_F)$ based on Eq. (1). The TTL is calculated by DFT procedure in terms of following equation.

$$\varepsilon(q/q') = \frac{E_D(q) - E_D(q')}{q'-q} \qquad (2)$$

The details form has been similar discussed by Janotti et al[21] and Zunger et al[22-24]. It is advantageous to use the idea of transition level because it can be observed in experiments where the final charge state of the local structure is capable to fully relax towards its equilibrium state after the charge transition or thermal ionization, by deep-level transient spectroscopy (DLTS)[25].

Consider the case of defect with charge q≠0, it needs a coulomb potential correction to counteract the effect provided by its image charge from the crystal lattice. Up to date, there have been three corrections developed. One is the band alignment developed by Van de Walle and coworkers[26], the second is the dispersion corrections of defect between the gamma point and Monkhorst-Pack by Wei et al[27]. The third is the dipole corrections which has considered the Madelung effect based on the static electric coulomb potential corrections, by Makov and Payne[28], which is the current popular corrections in DFT and used by us in this work. However, such image charge corrections contribute only around ~0.2 eV magnitude for the defect formation energy.

By this technique, we hold the opinion that, the O-interstitial defect in fact has the same chemical potential trend as the Cu-vacancy, and is the result of the deep hole trap level. Furthermore, the total energy of monoclinic CuO with corrected Hubbard U is difficult to determine because it provides the lower limit of the Cu chemical potential for the defect formation energy calculations in $Cu_2O$.

FIG. 2 presents the variation behaviors of d and p orbitals for the $d^{10}$-based compounds. The p orbitals perturbed by linear response also have a crossover behavior similar to that of the $d^{10}$ orbitals. The strong *p-d* coupling leads to a large portion of charge transfer from adjacent of d orbitals to the p orbitals of valence electrons of the anion elements. This elucidates the validation for the total energy with related to the occupation number of electron system[29-33].

According to FIG. 2, we obtained a d-orbital Hubbard correction of 6.8 eV for Cu and normalized 12 eV for O 2p orbitals. The twice-large Hubbard correction for the 2p orbitals of O occurs because each O site experiences two Cu-localized electron perturbations in the linear response calculations. As shown in FIG. 2, Cu presents the fully filled shell feature in the $Cu_2O$ system as it touches the 0 eV level. However, each perturbation of the d-electrons shared with one O site. Therefore, this is different from other ordinary coordinated metal oxides because metal atoms typically have a larger coordination number when bound to O. Therefore, $Cu_2O$ has a reversed $CaF_2$ structure because O occupies the Ca site, whereas Cu occupies the F site. The band structures of $Cu_2O$ and CuO in the AFM phase are shown in FIG. 3 (a) and (b).

An extensive experimental measurement on the electronic structure of $Cu_2O$ has been down by Ghijsen et a[34]l, we would like to give a comparison here. Consider the error and spectrum broadening factors, we also choose a 0.5 eV broadening for our valence band total density of states (TDOS) for comparison. Their reported Cu 3d spectrum is concentrated at 0~4 eV below the highest occupied level (0 eV), and the O 2p spectrum around 6~7 eV. The 2p-3d hybridization is presented at 4~8 eV. While from our calculations, we present a physical trend that corresponds to the experiment very well. As shown in the Fig. 2 (c), our calculated O 2p bands were ranged from 5 to 8.6 eV below the valence band maximum (VBM, 0 eV), and the Cu 3d bands contribute the top of the valence band that from 0 to 4 eV below the VBM (0 eV). Meanwhile, our method on the electronic structure of $Cu_2O$ also shows a substantial improvement like the work done by other groups[35-38].

The experimentally determined acceptor-like trap states are two states that are $E_V$+0.25 and $E_V$+0.45 eV [15]. However, because the oxygen flux was increased in those experiments, one of the densities of the trap states increased, whereas the other decreased to a lower density. This suggests that the O-related intrinsic defects play a significant role in providing acceptor-like trap levels near the VBM, and form unfavorable O-O homopolar bond under high O-concentration. However, as stated previously, this requires an accurate description of the localized hole states induced by the O-2p orbitals [2].

As shown in Table 1, our method produces consistent lattice parameters and electronic band gaps of $Cu_2O$, which is experimentally determined to be 2.17 eV. Isseroff et al. consider whether the $V_{Cu}^{split}$ has a lower formation energy than the simple Cu vacancy. Our data are consistent with Scanlon et al. because the neutral $V_{Cu}^{split}$ is approximately 0.8 eV lower than the $V_{Cu}^{simple}$ with help of local lattice reconstructions. This indicates that the corrected O-2p orbital energies improve the defect formation studies. One may argue that the Hubbard U parameter correction for O-2p orbitals is too high, with a magnitude of 12 eV. However, the formation energies of O-interstitial defects in tetrahedral and octahedral ($I_O^{tet}$ and $I_O^{oct}$) have similar values as the data provided by HSE, with an HF interaction percentage of 0.275, in the work of Scanlon et al [16, 17]. Thus, the O-2p orbital correction, in terms of Hubbard U, does not affect either the defect formation energies or thermal dynamic transition levels in different charges.

As shown in FIG. 4 (a), the contribution of the p-type intrinsic conductivity of $Cu_2O$ does not originate from $V_{Cu}^{split}$. The formation energies of the $V_{Cu}^{split}$ and simple $V_{Cu}$ are similar to the work of Scanlon et al, and the $V_{Cu}^{split}$ is approximately 0.6 eV lower than the simple $V_{Cu}$. This confirms that the Cu atom favors a flexible structural relaxation to a more stable tetrahedral coordination. However, this is not the lowest-energy defect, and neither is the oxygen vacancy ($V_O$). Instead, the lowest-energy defect is the oxygen interstitials ($I_O$). The experiments show that $Cu_2O$ is stabilized in the high-temperature condition because it follows the following reaction process from CuO: $4CuO \rightarrow 2Cu_2O + O2\uparrow$. The $Cu_2O$ structure has many hollow channels through which oxygen can diffuse. The excess O is trapped in this channel and induces localized states to capture electrons. The tetrahedral $I_O$ ($I_O^{tet}$) has an even lower formation energy compared with $V_{Cu}^{split}$. But octahedral $I_O$ ($I_O^{oct}$) determines the upper bound of the Fermi level for extrinsic doping where causes the defect formation spontaneously. The upper doping limit energy is about $E_V$+1.8 eV referring to the 0 eV of formation energy that is nearly constant from O-poor to O-rich potential limits, shown in Figure 4 (a).

The single-particle levels shown in Figure 4 (b) demonstrate that $V_{Cu}^{split}$ provides deep localized hole trap states next to the conduction band minimum (CBM), which has a d-orbital feature. The simple $V_{Cu}$ has a localized state that is 0.4 eV higher than the VBM. $I_O^{tet}$ has two localized states that are 0.2 and 0.5 eV higher than the VBM. This is similar to the experimental observations, where the trap states have been reported to be 0.25 and 0.45 eV higher than the VBM [15]. The formation energy of $I_O^{tet}$ is approximately 0.3 eV lower than the $V_{Cu}^{split}$. The electrical transition levels of (-/0) for $V_{Cu}^{split}$ and $I_O^{tet}$ are $E_V$+0.08 and $E_V$+0.39 that corresponds to the intrinsic p-type behavior. Therefore, the defect formation energies and electronic properties are determined based on the well-counteracted self-interaction, which is induced by the spurious self-energy of the localized orbitals. The experimentally reported intrinsic p-type conduction was found to be contributed by both $V_{Cu}^{split}$ and the O-interstitial intrinsic defects in $Cu_2O$. Both have similar formation energies in the range from (0/-1) to (-1/-2) transition states, with two single-particle trapping levels higher than the VBM.

Finally, the description of localized hole levels using DFT is the complicated issue for metal oxides. For the metal vacancy or anion interstitial site, the induced holes (removal of electrons) are often localized at the p-π orbitals of nearby O-sites, which denote levels near the valence band maximum (VBM). To accurately calculate these hole-induced levels, the Hubbard U parameter is used to correct the O-2p orbital energies in metal oxides. These orbital energies have been proven to significantly improve their single-particle levels in the band gap [2, 39-41].

The author gratefully acknowledges the support of natural science foundation of China (NSFC) for Youth Scientist (Grant No. NSFC 11504309).

**Table 1.** Summary of the lattice parameters, formation enthalpies of $Cu_2O$ and CuO (AFM), defect formation energies in (Cu-rich/O-poor), single-particle levels, and transition levels.

| | | PBE[19] | GGA [20] | PBE+U [42] | PBE+U [18] | HSE(0.275) [18] | HSE(0.275) [16] | This work | Exp[43] |
|---|---|---|---|---|---|---|---|---|---|
| | Lattice (Å) | 4.31 | 4.32 | | | | | 4.28 | 4.27 |
| | Eg | 0.43 | 0.46 | 0.4 | | 2.12 | 2.12 | 2.16 | 2.17 |
| $\Delta H_f$ | $Cu_2O$ | | -1.24 | | -1.64 | -1.62 | -1.59 | -1.79 | -1.75 |
| | CuO | | | | -1.21 | -1.44 | -1.46 | -1.74 | -1.63 |
| $\Delta Hf$ (D, 0) (eV) | $V_{Cu}^{simple}$ | 0.70 | 0.47 | 0.41 | 1.10 | 1.34 | 1.15 | 3.20 | |
| | $V_{Cu}^{split}$ | 1.00 | 0.78 | 0.47 | 1.28 | 1.58 | 1.14 | 2.48 | |
| | $I_O^{oct}$ | 1.80 | 1.90 | | | | 1.94 | 2.12 | |
| | $I_O^{tet}$ | 1.30 | 1.47 | | | | 1.87 | 2.19 | |
| | $V_O$ | 0.80 | 0.90 | | | | 1.20 | 2.62 | |
| Single-particle level (eV) | $V_{Cu}^{simple}$ | 0.00 | 0 | 0.00 | | | 0.52 | 0.11 | |
| | $V_{Cu}^{split}$ | | | | | 0.57 | 1.12 | 0.16 | |
| | $I_O^{oct}$ | | | | | | 1.14 | 0.31 | |
| | $I_O^{tet}$ | | | | | | 1.05 | 0.41 | |
| | $V_O$ | | | | | | | 0.44 | |
| (-1/0) (eV) | $V_{Cu}^{simple}$ | 0.28 | 0.18 | | | | 0.23 | -0.89 | |
| | $V_{Cu}^{split}$ | 0.29 | 0.20 | | | | 0.47 | 0.08 | |
| | $I_O^{oct}$ | 0.66 | 0.45 | | | | 1.08 | 0.55 | |
| | $I_O^{tet}$ | 0.78 | 0.65 | | | | 1.27 | 0.39 | |

**FIG. 1**

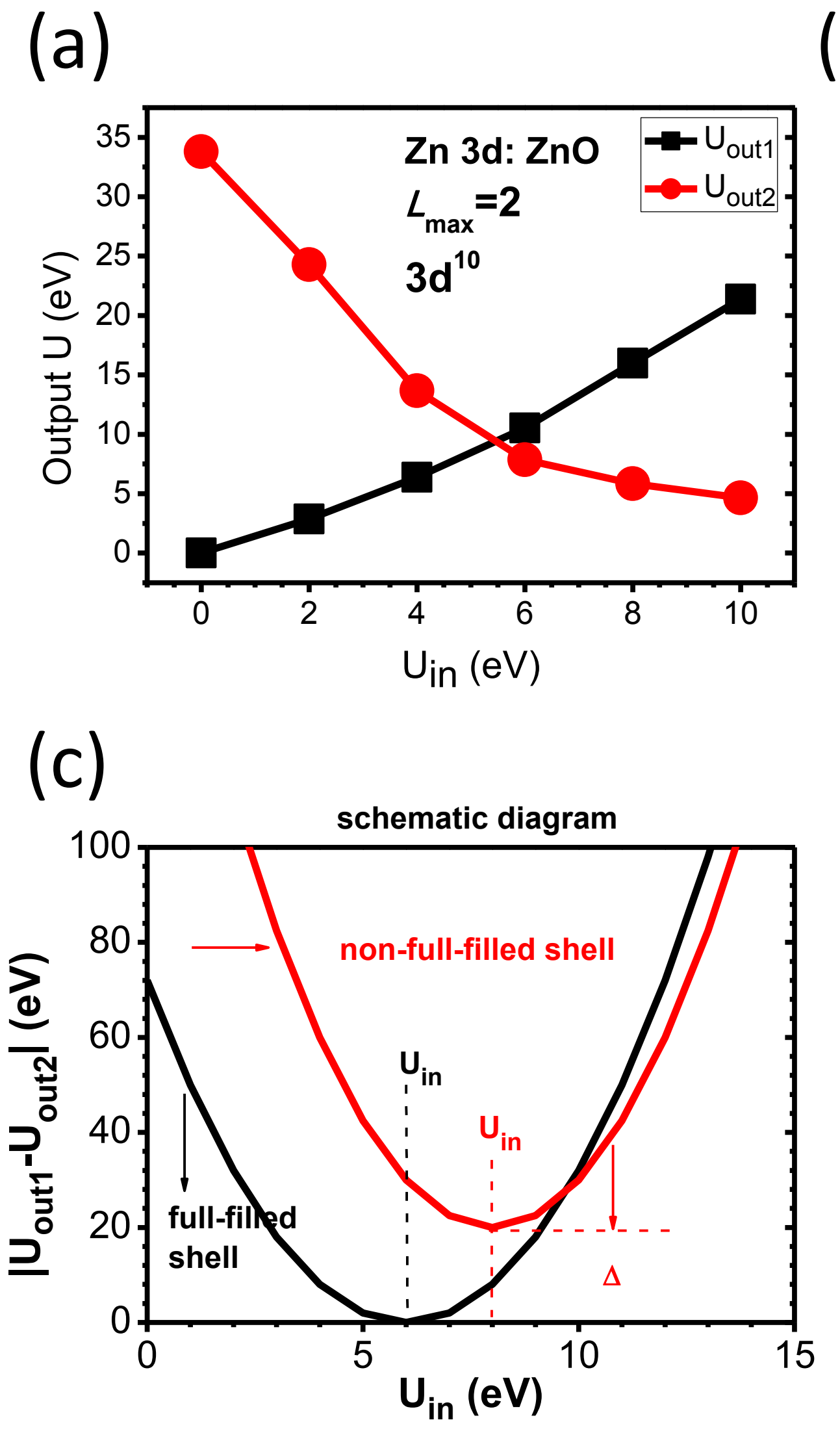

**FIG. 1**. Self-consistently obtained $U_{out1}$ and $U_{out2}$ for fully occupied orbitals from (a) ZnO, (b) $Cu_2O$. The cross-over feature of fully occupied shell denotes the $|U_{out1}-U_{out2}|=0$, shown in (c).

**FIG. 2**

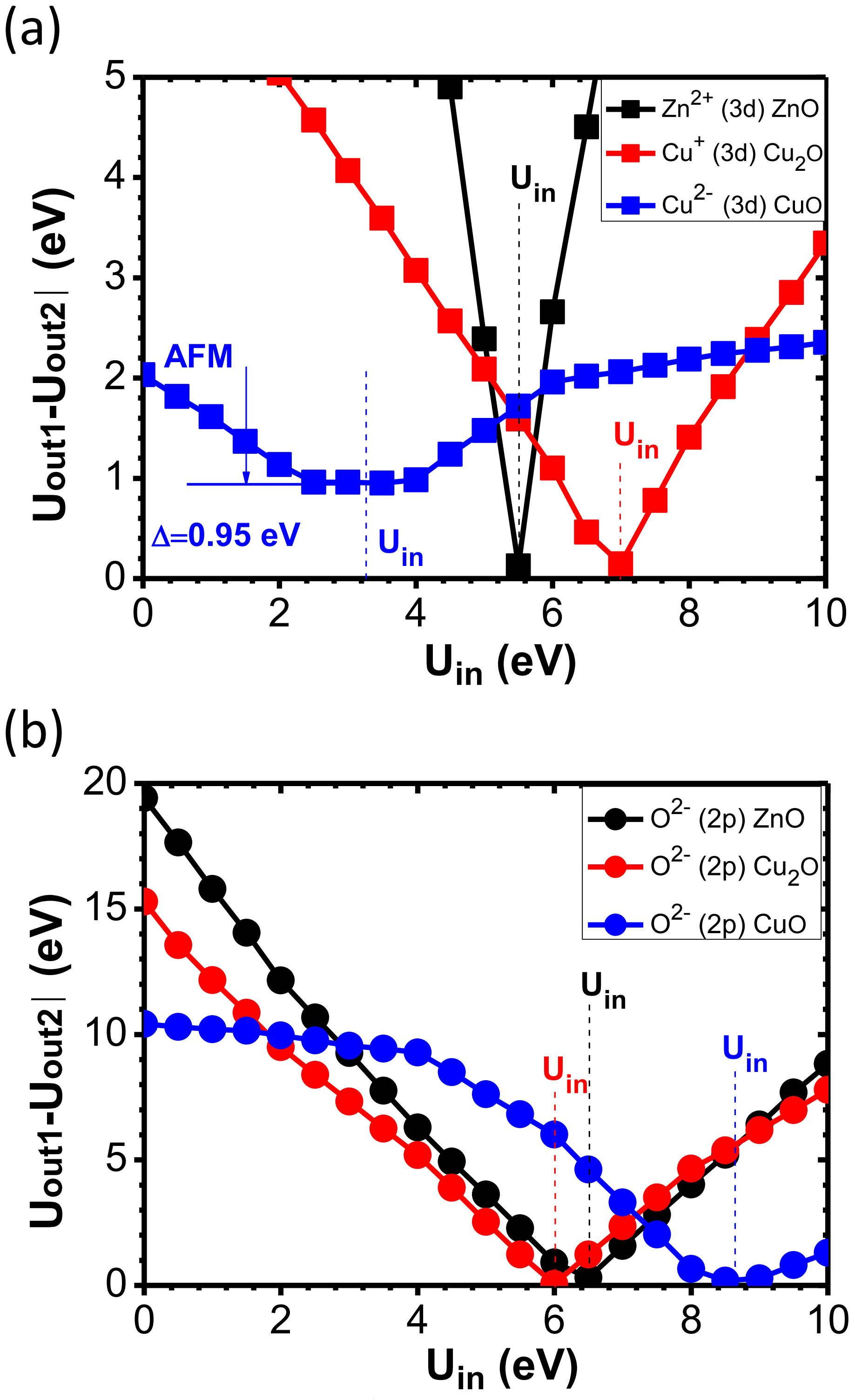

**FIG. 2**. The $|U_{out1}-U_{out2}|$ vs. $U_{in}$ behaviors of bulk wurtzite ZnO, bulk CuO in the AFM phase and bulk $Cu_2O$ structures with (a) d and (b) p localized electronic orbitals. (AFM: anti-ferromagnetic). $Cu^{2+}$ in CuO cannot achieve the full filled shell with a rigid Δ shift (a).

**FIG. 3**

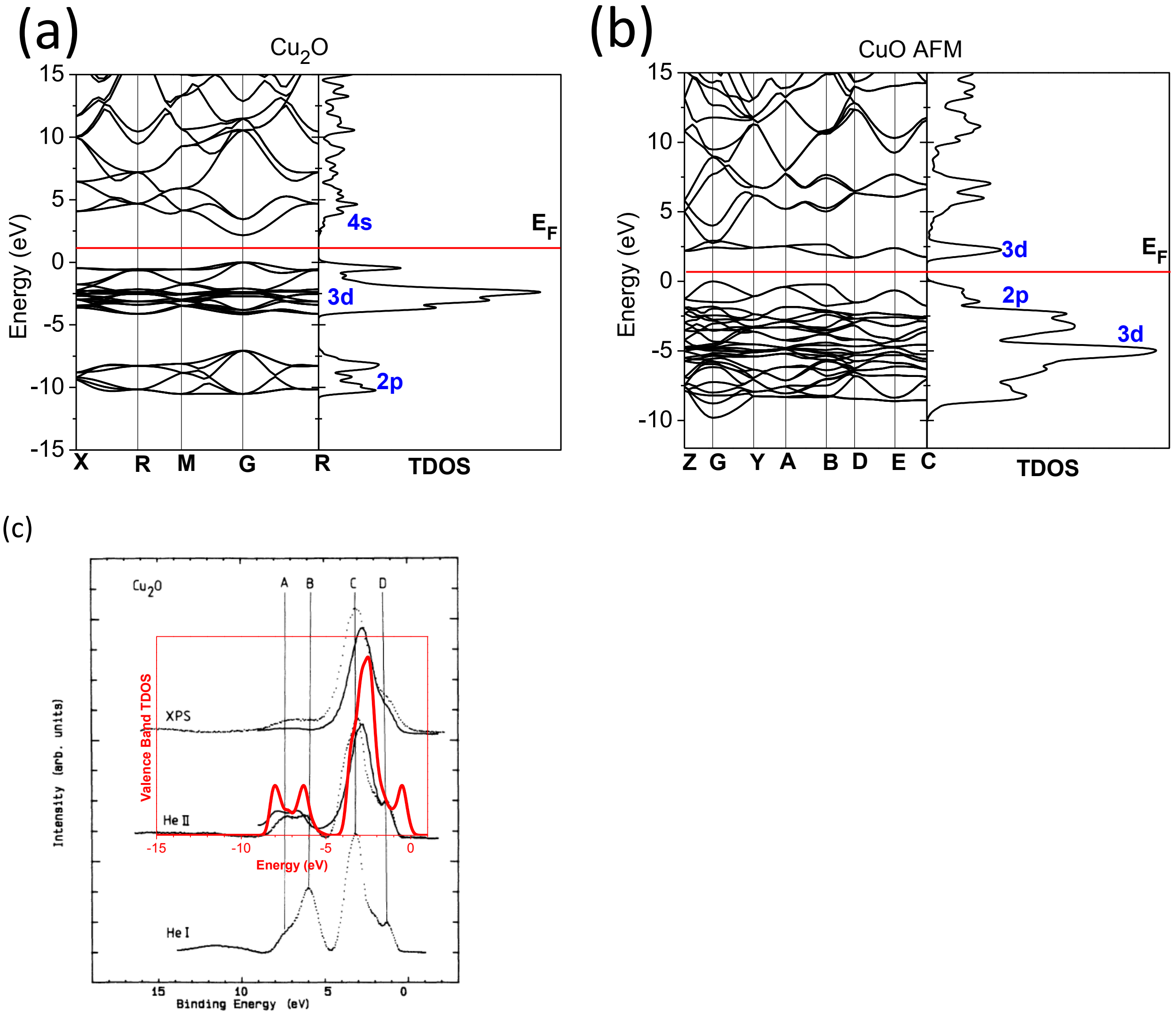

**FIG. 3.** (a) The band structure and TDOS of (a) $Cu_2O$ and (b) CuO in the AFM phase. (c) The comparison between our calculation (Red spectrum) and experimental measurements done by Ghijsen et al[34], the broadening factor of DFT spectra we chose is Gaussian-type with 0.5 eV.

**FIG. 4**

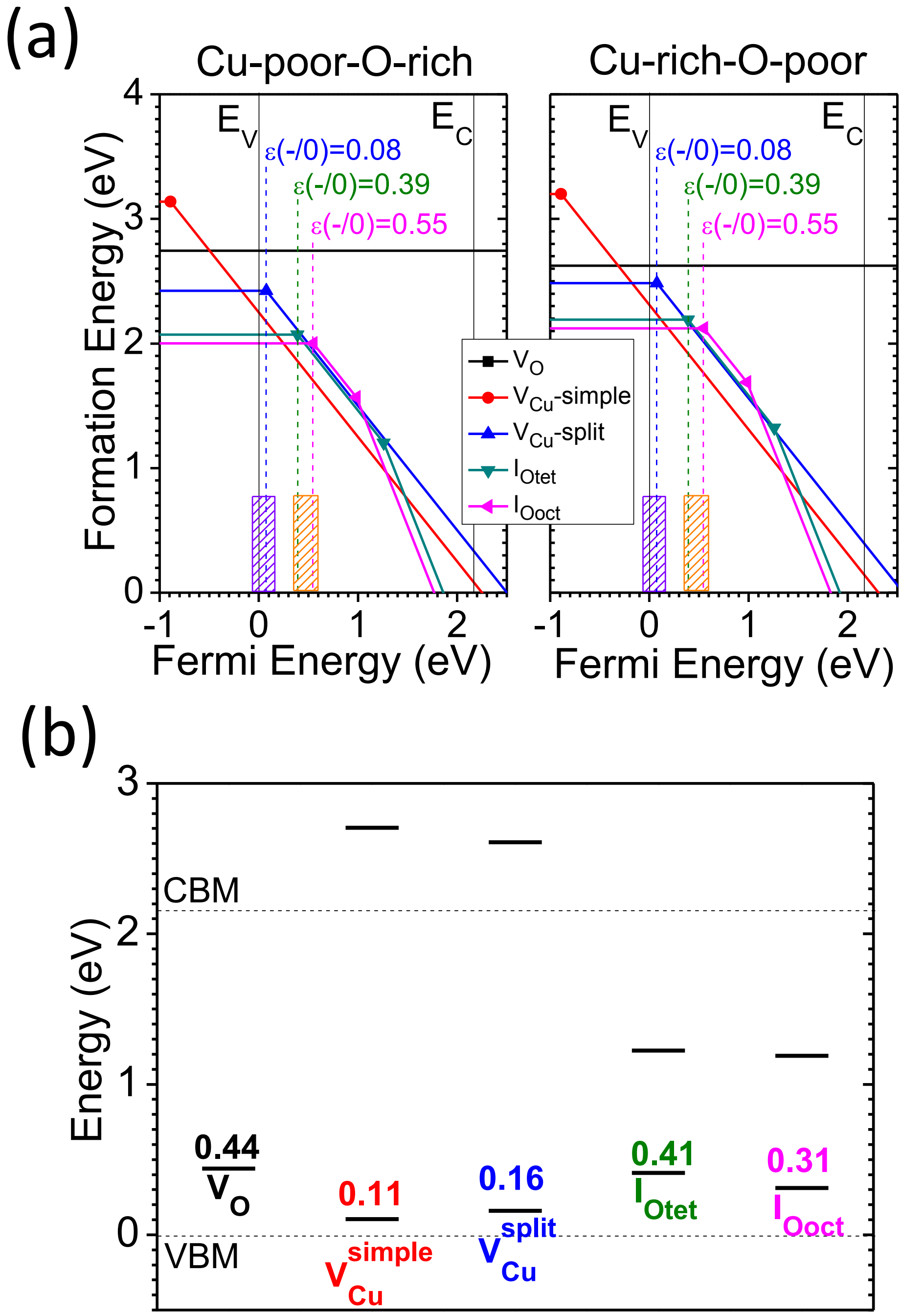

**FIG. 4.** (a) A summary of the formation energies of the intrinsic defects in $Cu_2O$ under O-rich and O-poor limit. (b) The localized single-particle level within the band gap of $Cu_2O$.